\documentclass[12pt]{article}
\usepackage{amsmath}
\usepackage{amsfonts}
\def\Box{\hbox{$\sqcup$\kern-0.66em\lower0.03ex\hbox{$\sqcap$}}}

\begin{document}
\begin{titlepage}
\begin{flushright}
IFUP--TH 2002/35 \\
\end{flushright}
\vskip 1truecm
\begin{center}
\Large{\bf Functional approach to $2+1$ dimensional gravity coupled to
particles}  
\footnote{This work is  supported in part by M.I.U.R.}
\end{center}
\bigskip\bigskip
\vskip 1truecm
\begin{center}
{Luigi Cantini$~^{a}$, Pietro Menotti$~^{b}$}\\  
\vskip 1truecm
{\small\it $^{a}$ Scuola Normale Superiore, 56100 Pisa, Italy\\
{\small\it and INFN, Sezione di Pisa}\\
e-mail: cantini@df.unipi.it\\
{\small\it $^{b}$ Dipartimento di Fisica dell'Universit{\`a}, 56100 Pisa, 
Italy}\\
{\small\it and INFN, Sezione di Pisa}\\
e-mail: menotti@df.unipi.it\\}
\bigskip\bigskip\bigskip

September 2002
\vskip 1truecm
\end{center}
\end{titlepage}

\begin{abstract} 
The quantum gravity problem of ${\cal N}$ point particles interacting
with the gravitational field in $2+1$ dimensions is approached working
out the phase-space functional integral. The maximally slicing gauge
is adopted for a non compact open universe with the topology of the
plane. The conjugate momenta to the gravitational field are related to
a class of meromorphic quadratic differentials. The boundary term for the
non compact space is worked out in detail. In the extraction of the
physical degrees of freedom functional determinants related to the
puncture formulation of string theory occur and cancel out in the
final reduction. Finally the ordering problem in the definition of the
functional integral is discussed. 
\end{abstract} 

\section{Introduction}\label{Introduction}
In this paper we examine the functional approach to a problem of
quantum gravity i.e. the quantum treatment of ${\cal N}$ particles
interacting with the gravitational field in $2+1$ dimensions. The analogous
problem in absence of matter has been dealt with by Carlip
\cite {carlip}. It is well known that in absence of matter gravity in
$2+1$ dimensions acquires a non trivial dynamics only on closed
universes and here the physical degrees of freedom are encoded in the
moduli of the space sections. The hamiltonian treatment of such a
problem is found in \cite{moncrief,hosoya,carlip2}; for genus $1$ i.e. torus
topology the classical hamiltonian is explicitly known and its
quantum transcription, choosing a proper ordering, gives rise to the
Maass laplacian, thoroughly studied by mathematicians
\cite{fayterras}. For higher 
genus no explicit expression is known even for the classical
hamiltonian.

In \cite{carlip} Carlip starts from the general form of the
phase space functional integral and through a process of gauge fixing
reduces it to a functional phase space integral on the physical
degrees of freedom i.e. the moduli and their conjugate momenta. The
final result is what one would have naively obtained by writing down
the simple minded phase space integral using the reduced action. This
is not obtained through particular tricks but simply interpreting the
functional $\delta$ functions in such a way as to preserve invariance
under diffeomorphisms. 

In presence of particles the problem acquires a highly non trivial
dynamics also on open spaces. For the open space with the
topology of the plane the maximally slicing gauge can be adopted,
leading to notable simplifications \cite{BCVW,MS,CMS1}. The two particle
case can be solved exactly both at the classical and quantum level
while in presence of three or more particles the hamiltonian even if
perfectly defined \cite{CMS1,CMS3} cannot be written in explicit form. 

In the following we deal with the functional formulation of quantum
$2+1$ dimensional gravity coupled to particles in an open universe
with the topology of the plane. There are two noteworthy differences
with respect to the problem dealt with in \cite{carlip}. a) We are in
presence of a Riemann surface with punctures at the location of the
particles. b) The boundary terms play an essential role in the
dynamics of the problem.

With regard to point (a) the main difference is that the transverse
traceless part of the space cotangent to the space of the spacial
metrics is described by a class of meromorphic quadratic differentials
which turn out to be parametrized by the canonical momenta of the
particles.

With regard to point (b) we treat the non compact case as
suggested in \cite{hawkinghunter} i.e. to consider first the problem
with a fixed boundary and then to take the limit when the boundary
goes to infinity. In such a limiting process the values at the
boundary of the fields have to be chosen as to provide a regular non
trivial dynamics.

As always the central problem with the functional integration in
presence of gauge symmetries, like quantum gravity, is to extract the
relevant degrees of freedom from the gauge degrees of freedom. This
can be performed by introducing gauge fixings and evaluating the
ensuing Faddeev- Popov determinants, as is usually done, or by
applying the so called geometric approach
\cite{alvarez,polchinski,mottola} in  
which the gauge volume is simply factorized. Both procedures will be
applied here and as expected they show up to be equivalent. The
geometric approach has the aesthetic advantage to extract directly the
result without introducing two gauge fixings whose explicit form at
the end is completely irrelevant. In the extraction of the relevant
degrees of freedom several non trivial functional determinants are
produced; this are analogous to the determinants occurring in the
puncture formulation of string theory \cite{dhoker}. Despite the
complexity of the intermediate steps all such determinants except one
cancel out exactly. The last
remaining determinant can be reduced to 1 by a simple and natural
choice of the canonical variables.

The final expression of the functional integral is the reduced phase
space integral in which only the particle positions and momenta
occur. Unfortunately, as it happens also in the case examined in
\cite{carlip}, such a functional expression tells us little about
the ordering problem which is related to the final definition of the
functional integral. We shall discuss shortly this problem in the
last section.

The paper is organized as follows: In Section 2 after recalling the
structure of the classical action with boundary terms included, we
write down the phase space functional integral. Integration over the
lapse and shift functions provides the hamiltonian and diffeomorphism
constraints. In Section 3 we deal with the integration over the space
metric and the momenta conjugate to the space metric. Here the
geometric approach is most clear: one describes the general metric as
the result of the application of a space diffeomorphism to the metric
given in the conformal gauge, where the conformal factor is allowed
singularities at the punctures which are the particle positions. It
is useful to distinguish the diffeomorphisms which do not move the
punctures from those which move them.

Then one considers the variations of the metric which are square
integrable in the De Witt metric. An important point is that in order
to ensure square integrability of such variations we find that the
diffeomorphisms which describe the motion of the punctures have to be
correlated with the motion of the singularities of the conformal
factor. Then one examines the space cotangent to such variations
i.e. the space of the conjugate momenta to the metric. It turns out
that the transverse traceless part of such a space is described by the
meromorphic quadratic differentials having only simple poles on the
punctures. In addition square integrability, in the De Witt metric, of
such conjugate momenta imply a sum rule on the residues; such a
restriction was already found in the classical treatment along a
different path by Deser \cite{deser} as a
necessary restriction on the conjugate momenta if one wants to avoid
singularities in the space metric. In \cite{CMS1,MS1} it arose from
the consistency of the asymptotic behavior of the conformal factor
with the hamiltonian constraint. Here it appears as a necessary
restriction on the square integrability of the conjugate momenta and
their variations.

In Section 4 we deal with the gauge fixings; the main role is played
by the maximally slicing gauge fixing; it is in fact such a fixing
which allows the solution of the diffeomorphism constraint in terms of
the quadratic differentials and reduces the hamiltonian constraint to
an equation equivalent to the Liouville equation.  At the end of this
section we briefly describe the geometric approach to the space
diffeomorphisms which as announced is equivalent to the Faddeev-Popov
procedure. In Section 5 we give a detailed derivation of the effective
hamiltonian for the non compact space as the limit of the boundary
term, when the boundary goes to infinity. As already mentioned the
procedure is the one suggested in \cite{hawkinghunter} which in our
case can be taken to the end explicitly. Finally we give a discussion
of the ordering problem in relation to other approaches to the
quantization of $2+1$ dimensional gravity.

\section{The action and the gauge fixings}
The gravitational part of the  action in $D+1$ dimensions, in hamiltonian
form is given by \cite{hawkinghunter,hayward}
\begin{multline}\label{azione}
S^{Grav} = \int_{{\cal M}} dt d^Dx \left[ \pi^{ij}\dot h_{ij} -
N^i H_i^{Grav} - NH^{Grav}\right]+ \\ 
+2\int dt \int_{B_t} d^{(D-1)}x
\,\sqrt{\rho} N \left( K_{B_t}+\frac{\eta}{\cosh\eta}
\mathcal{D}_\alpha
v^\alpha\right) -2\int dt \int_{B_t} d^{(D-1)}x\, r_\alpha
\pi^{\alpha\beta}_{(\rho)} N_\beta 
\end{multline} 
where $h_{ij}$ is the $D$-dimensional metric of the constant time
slices $\Sigma_t$, $\sinh \eta=n_\mu u^\mu$ with $n^\mu$ the
future-pointing unit normal to the time slices $\Sigma_t$ and $u^\mu$
the outward-pointing unit normal to the space like boundary $B$,
$B_t= \Sigma_t\cap B$,
$\sqrt{\rho}$ is the volume form induced by the space metric on $B_t$,
$K_{B_t}$ is the extrinsic curvature of $B_t$ as a surface embedded in 
$\Sigma_t$, $v_\alpha=\frac{1}{\cosh \eta}(n_\alpha-\sinh \eta ~u_\alpha)$
and $r_\alpha$ is the versor normal to $B_t$ in $\Sigma_t$.
The subscript $\rho$ in $\pi^{\alpha\beta}_{(\rho)}$
is a reminder that it has to be considered as a tensor density with respect
to the measure $\sqrt{\rho}$. $\mathcal{D}_\alpha$ is the covariant 
derivative induced by the metric on the space-like boundary $B$.
For the ADM metric \cite{ADM} we use the notation
\begin{equation}
ds^2 = -N^2 dt^2 +h_{ij}(dx^i+N^i dt)(dx^j+N^j dt).
\end{equation}
Moreover we have
\begin{equation}
H^{Grav}_i = -2 \sqrt h D_j\frac{\pi^j_{~i}}{\sqrt{h}}
\end{equation}
\begin{equation}
H^{Grav} = \frac{1}{\sqrt{h}} h_{ij}h_{kl} (\pi^{ik}\pi^{jl}-\pi^{ij}\pi^{kl})
-\sqrt{h}R,
\end{equation}
where $D$ is the covariant derivative induced by the
metric $h_{ij}$ on the surfaces $\Sigma_t$ and $R$ is the intrinsic
curvature of these surfaces.

The matter part of the action is given by
\begin{equation}\label{azionemateria}
S_m = \int
dt \sum_n P_{ni}\dot q^i_n+N^i(q_n)P_{ni}-N(q_n)
\sqrt{P_{ni} P_{nj} h^{ij}(q_n)+ m_n^2} .
\end{equation} 

In the following we shall denote
\begin{equation}
H_i=H_i^{Grav}-\sum_n P_{ni}\delta^2(x-q_n)
\end{equation}
and
\begin{equation}
H_0=H=H^{Grav}+\sum_n \sqrt{P_{ni}P_{nj} h^{ij}(q_n)+m^2_n}~\delta^2(x-q_n)
\end{equation}
and $S=S^{Grav}+S_m$.

As discussed in detail in \cite{hawkinghunter,hayward} 
$S$ is the classical action when the fields are kept constant on the
boundary $B_t$ i.e. the variation of $S$ provides the correct
equations of motion when such a variation is performed by keeping
constant the $2+1$ dimensional metric $g_{\mu\nu}$ on the boundary
$B_t$. As we shall be interested in an open universe we must 
take at the end the limit when the boundary $B_t$ goes to
infinity. For doing that we shall need to give the asymptotic
behavior of the fields at the boundary $|z|=r_0$ for $r_0\rightarrow
\infty$. This limit process  will be dealt with in Section 5
following the procedure described in \cite{hawkinghunter}.

We write the phase space functional integral as
\begin{equation}\label{functional1}
Z= \int 
\prod_{n=1}^{\mathcal{N}}D[P_n]
D[\pi^{ij}] D[h_{ij}] D[N^i] D[N] e^{i S}.
\end{equation}
The metric $h_{ij}$ is defined on the punctured plane
$R^2\setminus\{q_1,\dots q_{\cal N}\}$, where $q_1,\dots q_{\cal N}$ are
the particle positions. Thus integration on the metric
$h_{ij}$ implicitly contains the integration on the particle
positions. We shall derive the explicit form of such a dependence in
Section 3.
The functional integral (\ref{functional1}) is ill defined due to the
invariance of the action under space-time diffeomorphisms and as well
known one has to introduce $D+1=3$ gauge fixings, which we shall denote
by $\displaystyle {\delta(\chi)\prod_{i=1}^2\delta(\chi^i)}$. In
presence of such gauge fixings the functional integral takes the
form, known as Faddeev formula \cite{henneaux}
\begin{equation}
Z= \int 
\prod_{n=1}^{\mathcal{N}}D[P_n]
D[\pi^{ij}] D[h_{ij}] D[N^i] D[N] \delta(\chi)\prod_{i=1}^2\delta(\chi^i)
|{\rm Det}\{\chi^\mu, H_\nu\}| e^{i S}.
\end{equation}
$|{\rm Det}\{\chi^\mu, H_\nu\}|$ is the jacobian which assures the invariance 
under diffeomorphisms i.e. the Faddeev-Popov determinant.
We now integrate over the Lagrange multipliers $N^i$ and
$N$ obtaining apart for a multiplicative constant
\begin{equation}
Z= \int 
\prod_{n=1}^{\mathcal{N}}D[P_n]
D[\pi^{ij}] D[h_{ij}] \delta(\chi)\delta(\chi^1)\delta(\chi^2)
|{\rm Det}\{\chi^\mu, H_\nu\}|
\delta(\frac{H_i}{\sqrt{h}})\delta(\frac{H}{\sqrt{h}}) 
e^{i S}.
\end{equation}
The integration over $D[N]$ gives rise to $\delta(\frac{H}{\sqrt{h}})$
due to the following reason (see also \cite{carlip}). We recall that $N$ is a
scalar on the hypersurface $\Sigma_t$ (time-slice) and $H$ is a scalar
density on the same hypersurface.  Then the diffeomorphism invariant functional
extension of the formula $\delta(x)= \frac{1}{2\pi}\int e^{ipx}dp$ for
a scalar $s$ is
$$
\delta(s)={\rm const.} \int D[N] e^{i\int Ns\sqrt{h}d^Dx}
$$
and thus
$$
 \int D[N] e^{i\int HNd^Dx}={\rm const.}~\delta(\frac{H}{\sqrt{h}})
$$
and similarly for the densities $H_i$.

\noindent

\section{Integration over $D[h_{ij}]$ and
$D[\pi^{ij}]$}\label{IntegrationDhDpi}

In order to perform the integration in $D[h_{ij}]$ and $D[\pi^{ij}]$
we have to study the space of the metrics over the punctured plane
 and its cotangent space \cite{bers,dhoker}.  The general
parameterization of these metrics is
\begin{equation}
h_{ij} = F^*( e^{2\sigma} \delta_{ij})
\end{equation}
being $F$ a 2-dimensional diffeomorphism.
In 
fact for the punctured 2-dimensional plane the only Teichm\"uller
parameters are the positions of the punctures \cite{bers} and thus all
the metric through a diffeomorphism can be brought to the conformal
type.  $e^{2\sigma}$ will be a smooth conformal factor on the
punctured plane. Finite geodesic distance among the punctures allows
singularities 
\begin{equation}\label{singularities}
[(z-z^c_n)(\bar z-\bar z^c_n)]^{-\mu_n}~~~~{\rm~with}~~~~\mu_n<1,
\end{equation}
where we have employed the complex coordinate $z=x+iy$. We denote with
$z^c_n$ the position of the particles in the conformal gauge
$z^c_n=F(z_n)$.

As usual in the ADM approach \cite{ADM} we have to fix the
boundary conditions on the fields; we shall assume on a large circle
of radius $R$ the space metric diffeomorphic to ${\rm const}\times(z\bar
z)^{-\mu_0}\delta_{ij}$ with $0<\mu_0<1$. In the variational problem we
have to keep the metric at the boundary fixed or better the variation
of the metric $g_{\mu\nu}$ has to be such as not to vary the metric
induced on the boundary \cite{wald}.  So we can write
$e^{2\sigma}=e^{2\sigma_R}e^{2\sigma_S}$, where $\sigma_R$ is a
regular conformal factor, and $\sigma_S$ is given by
\begin{equation}
2\sigma_S =\sum_n -\mu_n \rho(z-z^c_n)\ln|z-z^c_n|^2 - 
\mu_0~\rho(\frac{1}{z})\ln(z\bar z),
\end{equation}
with $\rho(z)$ a smooth function having support inside a
circle of radius $1$.

For the functional integration in $D[h_{ij}]$ as done in
\cite{carlip,alvarez,polchinski} we
assume the measure induced by the diffeomorphism invariant distance provided by
the De Witt metric
\begin{equation}
(\delta h_{ij}, \delta h_{ij}) = \int\sqrt{h}~ \delta h_{ij}
G^{iji'j'}\delta h_{i'j'}d^2x
\end{equation}
with
\begin{equation}
2G^{iji'j'}=(h^{ii'}h^{jj'}+h^{ij'}h^{ji'}-\frac{2}{D}h^{ij}h^{i'j'})+
C~h^{ij}h^{i'j'},
\end{equation}
that is we set
\begin{equation}
1=\int D[\delta h_{ij}] e^{-(\delta h_{ij}, \delta h_{ij})}.
\end{equation}

In order to have a positive definite metric we need $C>0$; it is
however well known that the integration on the Weyl deformations of the
metric makes $C$ to disappear from the final result
\cite{polyakov,menottipeirano}.

If we want the measure $D[h_{ij}]$ to be well defined we must admit
only variations of the metric with finite norm
$(\delta h_{ij}, \delta h_{ij}) < \infty$.
This condition imposes the behavior at infinity
\begin{equation}\label{condinfty}
|\delta h_{ij}| \simeq (z\bar z)^{-\frac{\mu_0+1}{2}-\epsilon},~~~~
\epsilon >0.
\end{equation}
Similarly the behavior at the punctures $z^c_n$ must be
\begin{equation}\label{condpunct}
|\delta h_{zz}| \simeq (\zeta
 \bar\zeta)^{-\frac{\mu_n+1}{2}+\epsilon}~~~~{\rm with}~~~~\zeta=
 z-z^c_n. 
\end{equation}
We write now the variation of the metric. This is given by
\begin{equation}\label{decompdeltah}
\delta h_{ij} = \mathcal{L}_{\eta} h_{ij} + F^*((\delta
e^{2\sigma})\delta_{ij})
\end{equation}
with $\eta$ a vector field and $ \mathcal{L}_{\eta}$ the related Lie
derivative. For $F$ equal to the identity we have
\begin{equation}
\delta h_{ij} = \mathcal{L}_{\eta} (e^{2\sigma}\delta_{ij}) + \delta 
e^{2\sigma}\delta_{ij}.
\end{equation}

Due to the invariance of the De Witt metric under diffeomorphisms the above
reasoning and bounds extend also to the case $F\neq I$. In fact
\begin{equation}
\mathcal{L}_{\eta}
F^*(e^{2\sigma}\delta_{ij})=
F^*\mathcal{L}_{\xi}(e^{2\sigma}\delta_{ij})
\end{equation}
with $\xi = F_*\eta$.
Our purpose will be to change over from the integration on
$h_{ij}$, to the integration on the diffeomorphisms and the conformal
factor. Due to the presence of the gauge fixings the integration on the
diffeomorphisms explores only the infinitesimal ones i.e. the
tangent space described by vector fields $\xi$. 
With regard to the term
$ \mathcal{L}_{\xi} h_{ij}$ it will be instrumental to decompose the
field $\xi$ which generates the infinitesimal diffeomorphism as the sum
\begin{equation}\label{xidec}
\xi = \xi^{0} + \sum_{k=1}^{\cal N} \alpha_k \xi^{k}
+\sum_{k=1}^{\cal N} \bar \alpha_k \bar \xi^{k}
\end{equation}
with 
\begin{equation}\label{xipunct}
\bar \xi^{k\bar z}(z) = \overline {\xi^{kz}(z)},~~~
\xi^{kz}(z^c_n)=\bar\xi^{k\bar z}(z^c_n)=\delta_{kn},~~~
\xi^{k\bar z}(z)=0,~~~
\bar \xi^{kz}(z)=0
\end{equation}
and the field $\xi^{0}$ vanishes at the
punctures. The variation of the metric due to infinitesimal
diffeomorphisms becomes
\begin{equation}\label{dech}
\delta h_{ij} = (P\xi^{0})_{ij} + \sum_{k=1}^{\cal N} \alpha_k
(P\xi^{k})_{ij} + 
\sum_{k=1}^{\cal N} \bar \alpha_k (P\bar \xi^{k})_{ij} +
h_{ij}D_l\xi^l 
\end{equation}
with
\begin{equation}\label{defP}
(P\xi)_{ij} = D_i\xi_j+D_j\xi_i -D_l\xi^l h_{ij}.
\end{equation}
We shall come back to eq.(\ref{dech}) after eq.(\ref{Qbase2}).
Eq.(\ref{defP}) in the conformal metric and complex coordinates takes the
form \cite{alvarez}
\begin{equation}\label{defPcz}
(P\xi)_{zz} = 2
e^{2\sigma}\frac{\partial}{\partial 
z} (e^{-2\sigma}\xi_{z}) = e^{2\sigma}\frac{\partial}{\partial
z}\xi^{\bar z},
\end{equation}
\begin{equation}\label{defPcbz}
(P\xi)_{\bar z\bar z} = 2
e^{2\sigma}\frac{\partial}{\partial 
\bar z} (e^{-2\sigma}\xi_{\bar z}) = e^{2\sigma}\frac{\partial}{\partial
\bar z}\xi^{z},
\end{equation}
\begin{equation}
(P\xi)_{\bar zz}= (P\xi)_{z\bar z}=0.
\end{equation} 
The adjoint $P^+$ of the operator $P$ acting on the space of $\xi^0$ 
equipped with the diffeomorphism invariant metric 
\begin{equation}
(\xi, \xi)=\int \sqrt h~ \xi^i \xi^j h_{ij} \frac{i}{2} dz \wedge d\bar z.
\end{equation}
for $\delta h\in \mathcal{D}(P^+)$ is given by 
\begin{equation}
(P^+ \delta h)_z = -4 e^{-2\sigma}\frac{\partial}{\partial \bar z}
\delta h_{zz}
\end{equation}
and
\begin{equation}
(P^+ \delta h)_{\bar z} = 
-4e^{-2\sigma}\frac{\partial}{\partial z} \delta h_{\bar z\bar z}.
\end{equation}
The contribution to $\delta h_{ij}$ due to the variation of the
conformal factor is
\begin{equation}
(\delta e^{2\sigma})\delta_{ij}=\delta_{ij} e^{2\sigma}\delta(2\sigma_R)+
\delta_{ij} e^{2\sigma}\delta(2\sigma_S)
\end{equation}
where
\begin{equation}\label{de_sigmaS}
\delta(2\sigma_S)=\sum_n\mu_n\left(\frac{\delta z^c_n}{z-z^c_n}+
\frac{\delta \bar z^c_n}{\bar z-\bar z^c_n}\right)\rho(z-z^c_n)
+\sum_n\mu_n\ln|z-z^c_n|^2 \left(\frac{\partial \rho}{\partial z}\delta z^c_n
+\frac{\partial \rho}{\partial \bar z}\delta \bar z^c_n\right).
\end{equation}
The variation of $\delta \mu_n$ would give rise to $\delta\sigma_S$
which are square integrable and thus such variation can be
reabsorbed in $\delta \sigma_R$; in the following 
the hamiltonian constraint will fix $\mu_n=\frac{m_n}{4\pi}$.

We shall see now that the  
imposition that $\delta h_{ij}$ be square integrable in the De Witt
metric imposes a relation among the $\delta z^c_n$ and the $\xi^{k}$.

The variation $\delta 
h_{zz}=(P\xi)_{zz}=e^{2\sigma}\frac{\partial}{\partial z}
 \xi^{\bar
z}$ and $\delta h_{z\bar z} = \delta_{ij}
e^{2\sigma}\delta(2\sigma_R)$ are always square integrable for regular
$\xi^{\bar z}$ and $\delta(2\sigma_R)$, square integrable 
in their respective norms. In fact in order to have a finite norm, $\xi$
must behave at infinity as
\begin{equation} 
|\xi^{z}|\simeq(z\bar z)^{\mu_0-\frac{1}{2}-\epsilon}
\end{equation}
and so we have
\begin{equation}\label{compinfhzz}
\delta h_{zz} = e^{2\sigma}\frac{\partial}{\partial z} \xi^{\bar z} \simeq
(z\bar z)^{-1-\epsilon}
\end{equation}
satisfying  condition (\ref{condinfty}) at infinity and
\begin{equation}
(\delta h_{zz}, \delta h_{zz}) = 
\int \sqrt h~ e^{2\sigma}\frac{\partial}{\partial z} \xi^{\bar z}
h^{z\bar z} h^{z\bar z}
e^{2\sigma}\frac{\partial}{\partial \bar z} \xi^z \frac{i}{2}dz\wedge
d\bar z= 
\int \sqrt h ~{\rm regular}~ \frac{i}{2}dz\wedge d\bar z
\end{equation} 
is finite  because of the local finiteness of the area and
the behavior (\ref{compinfhzz}) at $\infty$. 
The analysis for $\delta (2\sigma_R)$ is even easier because 
$$
(h_{ij}\delta(2\sigma_R), h_{ij}\delta(2\sigma_R)) = 
\frac{C}{2}\int \sqrt h
~h_{ij}\delta(2\sigma_R) h_{lk}\delta(2\sigma_R) h^{ij}h^{lk}~
\frac{i}{2}dz\wedge d\bar z =  
$$
\begin{equation}
=2C\int \sqrt h~(\delta(2\sigma_R))^2 ~\frac{i}{2}dz\wedge d\bar z=
2C(\delta(2\sigma_R), 
\delta(2\sigma_R) ),
\end{equation}
and the finiteness of the norm of $\delta(2\sigma_R)$ implies directly
the finiteness of the norm of $h_{ij}\delta(2\sigma_R)$.

The problem arises with the contributions
$\delta (2 \sigma_S)\simeq \mu_n(\frac{\delta z^c_n}{z-z^c_n}+
\frac{\delta \bar z^c_n}{\bar z-\bar z^c_n})$ and $D_l\xi^l \simeq 
\partial_l\xi^l-\mu_n\left(\frac{\xi^z}{z-z^c_n}+\frac{\xi^{\bar z}}
{\bar z-\bar z^c_n}\right) $ which are not separately square integrable
as 
\begin{equation}
\int \sqrt h \frac{\mu_n^2}{(z-z^c_n)(\bar z -\bar
z^c_n)}~\frac{i}{2}dz\wedge d\bar z 
\end{equation}
is always divergent at the singularity $z=z^c_n$.
Thus in order to have an integrable $\delta h_{ij}$ we need
\begin{equation}
\alpha_n = \delta z^c_n,~~~~\bar\alpha_n = \delta \bar z^c_n.
\end{equation}

Our next job is to compute the functional jacobian in the transition
from the integration variable $h_{ij}$ to the $\sigma_R, z^c_n,\bar
z^c_n, 
\xi^0$. This is achieved with standards methods going over to the
tangent space \cite{alvarez}
$$
1=\int D[\delta h_{ij}]e^{-(\delta h_{ij}, \delta h_{ij} )}=
$$
\begin{equation}
=J_h \int \prod_{k=1}^{\cal N} dz^c_k d \bar z^c_k  D[\delta
\sigma_R]D[\xi^0] 
e^{-( \delta h_{ij}, \delta h_{ij} ) }.
\end{equation}
The decomposition (\ref{xidec}) seems to introduce an
arbitrariness in the further developments, as conditions (\ref{xipunct}) 
are very far from fixing the $\xi^{k}$ completely, but we shall
see that all our results will be independent on the
choice of such $\xi^{k}$ respecting condition (\ref{xipunct}).

Now
\begin{equation}
(\delta h_{ij}, \delta h_{ij}) = 
(P\xi^0+\alpha_k P\xi^{k}+\bar\alpha_k P\bar\xi^{k}, P\xi^0+\alpha_k
P\xi^{k}+ \bar\alpha_k P\bar\xi^{k})+ 
2C(\delta\sigma_R, \delta\sigma_R).
\end{equation}
In order to proceed further we characterize the orthogonal complement
to $P\xi^0$. This is given by the solutions of
\begin{equation}\label{picroce}
(P^+\delta h)_z = - 4e^{-2 \sigma}\frac{\partial \delta
h_{zz}}{\partial \bar z}=0 
\end{equation}
for $\delta h_{zz}\in D(P^+)$. From expression (\ref{defPcz},
\ref{defPcbz}) 
of $P$ we see that in order that $\delta h_{zz}\in D(P^+)$ it is necessary
that $\delta h_{zz}=O(\zeta^\alpha)$ at the singularities with $\alpha
\geq -1$. On the other hand (\ref{picroce}) tells us that $\delta h_{zz}$
has to 
be analytic 
on the punctured plane i.e. meromorphic on the plane while $\alpha
\geq -1$
excludes double poles and thus $\delta h \bot P\xi^0$ is of the form
\begin{equation}
\delta h_{z z} =\sum_{k=1}^{\cal N}\frac{\lambda_k}{z-
z^c_k};~~~~\delta h_{\bar z \bar z} =\sum_{k=1}^{\cal N}\frac{\bar
\lambda_k}{\bar z- \bar z^c_k}. 
\end{equation} 
Square integrability at infinity imposes
$\displaystyle{\sum_{k=1}^{\cal N}\lambda_k =0}$ 
and thus a complete basis of square integrable
holomorphic quadratic differentials is given by
\begin{equation}\label{Qbase1}
Q_{kzz}=\frac{1}{z-z^c_k}-\frac{1}{ z-z^c_1},
~~~~Q_{k\bar z \bar z}=0,~~~~(k=2,\dots {\cal N}).
\end{equation}
Taking the complex conjugate we have also
\begin{equation}\label{Qbase2}
\bar Q_{kzz}=0,~~~~\bar Q_{k\bar z \bar z}=\frac{1}{\bar z-\bar
z^c_k}-\frac{1}{\bar z-\bar z^c_1}~~~~(k=2,\dots {\cal N}).
\end{equation}
We come back to eq.(\ref{dech}). We saw that the orthogonal complement
to $P(\xi^0)$ in the space of the traceless square integrable $\delta
h_{ij}$ is ${\cal N}-1$ dimensional. Thus if we leave in
eq.(\ref{dech}) the $\alpha_k$ arbitrary we introduce an overcounting.
This is due to the fact that for $\alpha_1=\alpha_2=\dots=\alpha_{\cal
N}$, $P\left(\sum_{k=1}^{\cal
N}(\alpha_k\xi^{k}+\bar\alpha_k\bar\xi^{k})\right)$ is 
orthogonal to all 
the meromorphic 
quadratic differentials $Q_k$ and $\bar Q_k$ and as such it belongs to
the closure of 
the space $P(\xi^0)$. Several choices, all
physically equivalent, can 
be done; we shall choose 
$\alpha_1=0$ which as we shall see shortly describes the dynamics in
the relative coordinates ${z'}^c_n=z^c_n-z^c_1$.  
Now we decompose $P\bar\xi^{k}$ into the two mutually orthogonal
contributions
\begin{equation}\label{decomposition}
(P\bar\xi^{k})_{zz} = (P\bar\xi^{0k})_{zz}+\sum_{m=2}^{\cal
N}\beta^k_m Q_{mzz},   
\end{equation}
with
\begin{equation}
\beta^k_m=\sum_{n=2}^{\cal N}(Q_m,\bar Q_n)^{-1}(\bar Q_n, P\bar\xi^{k})
\end{equation}
where $(Q_k,\bar Q_m)^{-1}$ denotes the inverse of the $({\cal
N}-1)\times ({\cal N}-1)$ matrix $(\bar
Q_m,Q_n)$ i.e. $\sum_{m=2}^{\cal N}(Q_k,\bar Q_m)^{-1}(\bar
Q_m,Q_n) = \delta_{kn}$ .  In the Appendix it is shown
that for a proper choice of the $\xi^k$, always satisfying condition
(\ref{xipunct}), such square integrable 
$\xi^{0k}$ exists.  
With the round brackets as always we understand the invariant scalar
product according to the metric $h_{ij}$. E.g.
\begin{equation}
(\bar Q_m, Q_n)=\int \sqrt h \bar
Q_{mij}~h^{ii'}~h^{jj'}~Q_{ni'j'}\frac{i}{2}dz \wedge d\bar z=
\int \sqrt{h} \bar Q_{m\bar z\bar z}~h^{\bar z z}~h^{\bar
z z}~Q_{nzz}\frac{i}{2}dz \wedge d\bar z.
\end{equation}
As a consequence $(Q_m, Q_n)=(\bar Q_m, \bar Q_n)=0$.

Thus performing the shift $\displaystyle{\xi^0\rightarrow 
\xi^0+\sum_{k=2}^{\cal N} (\delta {z'}^c_k \xi^{0k}}+\delta \bar{z'}^c_k
\bar\xi^{0k})$, in the integration  
over the space of the fields $\xi^0$ we have
\begin{equation}
1=J_h \int \prod_{k=2}^{\cal N} d{z'}^c_k d\bar {z'}^c_k
D[\delta \sigma_R]D[\xi^0] 
e^{-2C(\delta \sigma_R, \delta \sigma_R)} e^{-(P\xi^0, P\xi^0)}
e^{-2\delta {z'}^c_k \delta \bar {z'}^c_l(P\xi^{k}, Q_n)(Q_n,
\bar Q_s)^{-1}(\bar Q_s, P\bar \xi^{l})}
\end{equation}
and thus apart from irrelevant numerical factors we have
\begin{equation}
J_h = {\rm Det}^*(P^+P)^{\frac{1}{2}}~\det(\bar Q_k, Q_l)^{-1}
\det(Q_s, P\xi^{l})\det(P\bar \xi^{k},\bar Q_n)
\end{equation}
where ${\rm Det}^*(P^+P)$ is the determinant of the operator $P^+P$,
restricted to the space of the $\xi^0$ \cite{dhoker} and the
functional measure on 
the space of the $\xi^0$ is defined as
\begin{equation}
1=\int D[\xi^0]e^{-(\xi^0, \xi^0)},
\end{equation}
and that on the space of $\delta \sigma_R$ 
\begin{equation}
1=\int D[\delta \sigma_R] e^{-2C(\delta\sigma_R,\delta\sigma_R)}.
\end{equation}
Thus
\begin{equation}
D[h_{ij}]=D[F^0]\prod_{k=2}^{\cal N} d{z'}^c_k d\bar {z'}^c_k~
D[\sigma_R]{\rm Det}^*(P^+P)^{\frac{1}{2}}~ 
\det(Q_l, \bar Q_k)^{-1}~ \det(Q_s, \mu_l)~ \det(\bar \mu_l, \bar Q_s)
\end{equation}
with $\mu_l$ the Beltrami differential
\begin{equation}
\mu_l = P\xi^{l}.
\end{equation}

Now we pass to the integration on the $\pi^{ij}$, that is on the
space cotangent to the space of punctured metrics. 
We give a useful parameterization of
$\pi^{ij}$ in term of which to perform the integration, provided we
compute the functional determinant related to the change of
parameterization.  The measure $D[\pi^{ij}]$ is defined by
\begin{equation}
1=\int D[\pi^{ij}] e^{-(\pi^{ij},\pi^{ij})}
\end{equation}
where 
\begin{equation}\label{normapi}
(\pi^{ij}, \pi^{ij})=\int \frac{1}{\sqrt h}  \pi^{ij} G_{ijmn}\pi^{mn} 
\frac{i}{2}dz\wedge d\bar z.
\end{equation}
As in the case of the integration over the variation of the metric
$\delta h_{ij}$, we must impose the square integrability of $\pi^{ij}$
in the norm (\ref{normapi}) in order to have a well defined functional
measure.  Moreover we want to have a well defined action, so $\delta
h_{ij}$ and $\pi^{ij}$ must satisfy the following condition
\begin{equation}
\int \pi^{ij} \delta h_{ij} \frac{i}{2} dz \wedge d\bar z < \infty.
\end{equation}
Taking into account the restrictions previously imposed on $\delta
h_{ij}$ at the singularities and at infinity we obtain
\begin{equation}
\left(\frac{\pi^{ij} }{\sqrt h}\right) \simeq (z \bar
z)^{\frac{3}{2}\mu_0 -\frac{1}{2}-\epsilon} ~~~~{\rm at}~\infty
\end{equation}
as at infinity
\begin{equation}
\delta h_{ij} \simeq (z \bar z)^{-\frac{\mu_0+1}{2} -\epsilon}
\end{equation}
and on each singularity
\begin{equation}
\left(\frac{\pi^{ij} }{\sqrt h}\right) \simeq 
(\zeta \bar \zeta)^{\frac{3}{2}\mu_n -\frac{1}{2}+\epsilon}
\end{equation}
as the behavior of $\delta h_{ij}$ is
\begin{equation}
\delta h_{ij} \simeq (\zeta \bar \zeta)^{-\frac{\mu_n+1}{2} +\epsilon}.
\end{equation}
It is easy to verify that these behaviors assure also the square
summability of $\pi^{ij}$.

We give now the orthogonal decomposition of $\pi^{ij}$ obeying to the
restrictions described above. We set
\begin{equation}
\pi^{ij} = \frac{\pi}{2} h^{ij} + \pi^{T ij}
\end{equation}
where $\pi = \pi^{ij}h_{ij}$ and $\pi^{T ij}$ is a
traceless tensor. Furthermore we decompose $\pi^{T ij}$ in a
transverse part and a remainder, i.e. we write
\begin{equation}
\frac{\pi^{Tij}}{\sqrt h} =\frac{\pi^{TTij} }{\sqrt h} + (PY^0)^{ij},
\end{equation}
where by definition $\pi^{TT}$ belongs to the orthogonal (traceless)
complement to $(PY^0)^{ij}$, being $Y^0$ the square integrable vector
fields vanishing at the punctures. The previously defined $\pi^{TT}$
are solutions (square integrable) of the equation
\begin{equation}\label{Pcroce0}
\left(P^+\frac{\pi^{TT}}{\sqrt h} \right)_j =0
\end{equation}
on the punctured plane. In the conformal metric we can rewrite 
eq.(\ref{Pcroce0}) as
\begin{equation}
\left(P^+\frac{\pi^{TT}}{\sqrt h} \right)_z =
-4 e^{-2\sigma}\frac{\partial}{\partial \bar z} \frac{\pi^{TT}_{zz}
}{\sqrt h} =0,
\end{equation}
%and thus
%\begin{equation}
%\partial_{\bar z} \pi^{TT\bar z}_{~~~z}=0,
%\end{equation}
i.e. $\displaystyle{\frac{\pi^{TT}_{~~zz}}{\sqrt h}}$ is a
meromorphic function. On the other hand square integrability forbids
poles of order higher than the first at the punctures $z^c_n$ and the
behavior $1/z$ at $\infty$. Thus in terms of the basis
(\ref{Qbase1},\ref{Qbase2}) 
we can write 
\begin{equation}
\frac{\pi^{TT}_{~~zz}}{\sqrt h} = -\frac{1}{4\pi}\sum_{k=2}^{\cal N} t_k
Q_{kzz},~~~~\frac{\pi^{TT}_{~~\bar z\bar z}}{\sqrt h} =
-\frac{1}{4\pi}\sum_{k=2}^{\cal N} \bar t_k \bar Q_{k\bar z\bar z}. 
\end{equation}

We come now to the integration measure $D[\pi^{ij}]$; we want to
express it in terms of the variables, $Y^0, t_k, \bar t_k,
\frac{\pi}{\sqrt h}$. The Jacobian relative to this change of
parameterization is given by 
\begin{equation}
1=J_{\pi}~\int D\left[\frac{\pi}{\sqrt h}\right]~ \prod_{k=2}^{\cal N}
dt_k d\bar t_k~ 
D[Y^0]~ e^{-(\pi^{ij}, \pi^{ij}) }.
\end{equation}
But 
$$
(\pi^{ij}, \pi^{ij}) = 2C\int \sqrt h \left(\frac{\pi}{\sqrt h}\right)^2
\frac{i}{2} dz\wedge d\bar z +
$$
\begin{equation}
+(PY^0, PY^0)+ \frac{1}{16\pi^2}\sum_{k,l} \bar t_k t_l\int 2 \sqrt{h}~\bar
Q_{k\bar z\bar z}~h^{\bar z z}h^{\bar z z}~ Q_{lzz} \frac{i}{2} dz\wedge
d\bar z
\end{equation}
from which 
\begin{equation}
J_\pi = {\rm Det}^*(P^+P)^{\frac{1}{2}}~\det(\bar Q_k, Q_l ),
\end{equation}
having used the normalization
\begin{equation}
1= \int D\left[\frac{\pi}{\sqrt h}\right]
e^{-2C(\frac{\pi}{\sqrt h},\frac{\pi}{\sqrt h})}.
\end{equation}
Putting together $J_h$ and $J_\pi$ we have
$$
J_h \times J_\pi = {\rm Det}^*(P^+P)^{\frac{1}{2}}~ \det(\bar Q_l,
Q_k)^{-1}~ \det(Q_s, \mu_n)\det(\bar \mu_n, \bar Q_s)~ {\rm
Det}^*(P^+P)^{\frac{1}{2}} \det(\bar Q_l, Q_k)=
$$
\begin{equation}
= {\rm Det}^*(P^+P)~\det(Q_s, \mu_n)\det(\bar \mu_n, \bar Q_s).
\end{equation}
The phase space functional integral now reads
$$
Z= \int\prod_{n=1}^{\cal N}D[P_n]
\prod_{n=2}^{\mathcal{N}}D[t_n]D[\bar t_n]D[\frac{\pi}{\sqrt
h}]D[Y^0]D[{z'}^c_n]D[\bar{z'}^c_n] 
D[\sigma_R]D[F^0]~{\rm Det}^*(P^+P)\times
$$
\begin{equation}\label{Z4} 
\det(Q_s, \mu_n)\det(\bar \mu_n,
\bar Q_s)\delta(\chi)\delta(\chi^1)\delta(\chi^2) |{\rm
Det}\{\chi^\mu, 
H_\nu\}| \delta(\frac{H_i}{\sqrt{h}})\delta(\frac{H}{\sqrt{h}}) e^{i
S}.
\end{equation}
We notice that with our choice of $\xi^{k}$ eq.(\ref{xipunct}) we
have $(Q_s, \mu_n)=-4\pi \delta_{sn}$ and thus the last two
determinants give the numerical constant $(-4\pi)^{2(N-1)}$.

\noindent

\section{Constraints and gauge fixings}\label{Delta}

As already mentioned in the introduction we shall adopt the maximally
slicing gauge $\delta(\chi)$ with $\chi=\frac{\pi}{\sqrt
h}=\frac{\pi^{ij}h_{ij}}{\sqrt h}$. This will be the only relevant
gauge fixing as the other two $\delta(\chi^1),~\delta(\chi^2)$ have
the simple role of locking the space diffeomorphisms. The results
are completely independent of the explicit form of $\chi^1$ and
$\chi^2$; one could also adopt the viewpoint of
\cite{alvarez,polchinski,mottola} i.e. not to put any
gauge fixing for the diffeomorphisms and simply factorize the infinite
volume of the integration on the diffeomorphisms outside the
functional integral, as we shall discuss at the end of this section.

We first consider the constraint
\begin{equation}
\frac{H_i}{\sqrt{h}}=-2 D_j\frac{\pi^j_{~i}}{\sqrt h} -
\sum_{n=1}^{\cal N}\frac{P_{ni}}{\sqrt h}\delta^2(z-z_n). 
\end{equation}
This can be written in terms of the
$\displaystyle{\frac{\pi_{zz}}{\sqrt h}}$ in the conformal metric as
$$
-4 e^{-2\sigma}\partial_{\bar z} (\frac{\pi_{zz}}{\sqrt
h})=
$$
\begin{equation}
\left((P^+P)Y^0\right)_z+ 
\sum_{n=2}^{\cal
N}~\left(\delta^2(z-z_n^c)-\delta^2(z-z_1^c)\right)~\frac{t_n}{\sqrt h}=F_*
\left(\sum_{n=1}^{\cal N} \frac{P_n}{\sqrt h}\delta^2(z-z_n)\right)_z.
\end{equation}
Integration over $Y^0$ gives $Y^0=0$ and produces the diffeomorphism
invariant determinant $[{\rm Det}^*(P^+P)]^{-1}$ which will cancel the
one appearing in eq.(\ref{Z4}), while integration over $P_n$
produces
\begin{equation}
t_n = (F_* P_n)_z~~~~~(n=2\dots {\cal N}),~~~~\sum_{n=2}^{\cal N} t_n
= -\left(F_*P_1\right)_z
\end{equation}
\begin{equation}\label{Pntn}
\bar t_n = (F_* P_n)_{\bar z}~~~~~(n=2\dots {\cal
N}),~~~~\sum_{n=2}^{\cal N} \bar t_n = -\left(F_*P_1\right)_{\bar z}.  
\end{equation}
The phase space path integral now reads
\begin{equation}\label{Z5} 
Z= \int 
\prod_{n=2}^{\cal N}D[t_n] D[\bar t_n]D[{z'}^c_n] D[\bar {z'}^c_n] 
D[\sigma_R]~D[F^0]~
\delta(\chi^1)\delta(\chi^2)
|{\rm Det}\{\chi^\mu, H_\nu\}|
\delta(\frac{H}{\sqrt{h}}) 
e^{i S}.
\end{equation}
As noticed in \cite{carlip} the gauge fixing $\frac{\pi}{\sqrt h}=0$
has zero Poisson bracket with $H_i$,  being $H_i$ the generators of
the diffeomorphisms and thus
\begin{equation}
{\rm Det}\{\chi^{\mu}, H_{\nu}\}={\rm Det}\{\chi^i, H_j\}{\rm
Det}\{\chi, H\}. 
\end{equation}
We notice
\begin{equation}\label{volume_chi_i}
\int D[F^0] |{\rm Det}\{\chi^i, H_j\}|\delta(\chi^1)\delta(\chi^2)=
{\rm const.} 
\end{equation}
The functional integral becomes
\begin{equation}\label{functional_sigma} 
Z= \int \prod_{n=2}^{\cal N}D[t_n]D[\bar t_n]D[{z'}^c_n]D[{\bar {z'}}^c_n]
D[\sigma_R]~%|\det(Q_s, \mu_n)|
|{\rm Det}\{\chi, H\}|
\delta(\frac{H}{\sqrt{h}}) 
e^{i S},
\end{equation}
which is explicitly independent of the choice of the gauge fixings
$\chi^1$, $\chi^2$.
We give now the explicit form of the hamiltonian constraint
$$
0=\frac{H}{\sqrt h}=\frac{1}{h}\pi^{ab}\pi_{ab}-R+
\sum_n \frac{1}{\sqrt h}\delta^2(z-z_n) \sqrt{m_n^2+P_{ni}h^{ij}P_{nj}}=
$$
\begin{equation}\label{constrH}
=F^*\left(~e^{-2\sigma}\left(e^{-2\sigma}2\pi^z_{~\bar z}\pi^{\bar
z}_{~z}+\Delta_0(2\sigma)+ \sum_n \delta^2(z-z^c_n)
\sqrt{m_n^2+ 2e^{-2\sigma(z^c_n)} t_n \bar t_n}\right)~\right),
\end{equation}
where we took into account that $\pi=0$.
Eq.(\ref{constrH}) is satisfied by the solution of the equation \cite{MS}
\begin{equation}\label{liouville}
\Delta_0 2\sigma= - 2\pi^z_{\bar z} \pi^{\bar z}_z e^{-2\sigma}-\sum_n m_n
\delta^2(z-z_n^c)
\end{equation}
or equivalently by the solution of the Liouville equation
\begin{equation}\label{tildeliouville}
\Delta_0{2\tilde{\sigma}}=-e^{-2\tilde{\sigma}}-\sum_n (m_n-4\pi)
\delta^2(z-z_n^c) -4\pi \sum_A \delta^2(z-z^c_A)
\end{equation}
being $z^c_A$ the position of the apparent singularities i.e. those
values for which
\begin{equation}
\sum_{n=2}^{\cal N}\frac{t_n}{z^c_A-z^c_n}-\frac{\sum_{n=2}^{\cal N}
t_n}{z^c_A-z^c_1}=0, 
\end{equation}
and $e^{-2\tilde{\sigma}}=e^{-2\sigma}2\pi^z_{~\bar z}\pi^{\bar
z}_{~z}$. As proved in \cite{picard, licht, troyanov} the
solution of eq.(\ref{liouville}) is unique for $e^{2\sigma}$ behaving
at infinity like $(z\bar z)^{-\mu}$ with $\mu<1$, $0<\frac{m_n}{4\pi}$
and $\sum_n \frac{m_n}{4\pi}<\mu$.  Substituting the behaviors
(\ref{singularities}) in eq.(\ref{liouville}) we see that the $\mu_i$
are fixed to the constant particle masses $\mu_n=\frac{m_n}{4\pi}$ and
the requirement of $2+1$ dimensional gravity $0<\mu_n<1,~0<\mu<1$
satisfy Picard  bound. We recall that $0<\mu_n<1$ states that the
$n$-th particle mass is positive and that the conical deficit at $q^c_n$ is 
less than $2\pi$; $0<\mu<1$ states that the total energy is positive
and that the total conical defect is less than $2\pi$, while
$\sum_n \mu_n <\mu$ states that the total energy exceed the sum of the rest
masses \cite{MS,CMS1}. 

Next we have to compute ${\rm Det}\{\chi,H\}={\rm
Det}\{\frac{\pi}{\sqrt h},H\}$ when the two conditions $\pi=0, H=0$
are satisfied. Such a calculation can be performed by using the
relations $\{A(x),\pi^{ij}(y)\}=\frac{\delta A(x)}{\delta h_{ij}(y)}$
and $\{A(x),h_{ij}(y)\}=-\frac{\delta A(x)}{\delta \pi^{ij}(y)}$. The
functional derivative with respect to $h_{ij}$ can be computed with
the help of the standard formula \cite{wald}
\begin{equation}
h^{ab}(x)\frac{\delta R_{ab}(x)}{\delta h_{ij}(y)}=
(D^iD^j-h^{ij}D^aD_a)\delta^2(x-y)
\end{equation}
obtaining for such a determinant 
\begin{equation}\label{detchiH}
{\rm Det}\{\chi,H\}={\rm
Det}\left[\left(\frac{1}{h}\pi^{ab}\pi_{ab}-D^2\right)\delta^2(x-y)\right]
\end{equation}
being $D^2=D^aD_a$ the scalar Laplace-Beltrami operator.

We show now that the determinant (\ref{detchiH}) cancels with the one
arising from $\delta(\frac{H}{\sqrt h})$. In fact from
eq.(\ref{constrH}) when we integrate $\delta(\frac{H}{\sqrt h})$ in
$D[\sigma_R]$ we obtain as a result
\begin{equation}
\left[{\rm Det}F^*(~e^{-2\sigma}(-e^{-2\sigma}2\pi^z_{~\bar z}\pi^{\bar
z}_{~z}+\Delta_0)\delta^2(x-y)~)\right]^{-1}
\end{equation}
where we took into account the constraint $\frac{H}{\sqrt h}=0$ and
the vanishing of the functional derivative of the source term, always
on $\frac{H}{\sqrt h}=0$, due to the behavior $e^{-2\sigma}\simeq{\rm
const.}(\zeta \bar{\zeta})^{\mu_n}$ at the particle singularities \cite{MS}. 

The action $S$ after solving the constraints becomes
\begin{equation}
\int dt \left[\sum_{n=1}^{\cal N} P_{ni}~\dot q^i_n + \int
\pi^{ij}\dot h_{ij} \frac{i}{2}dz\wedge d\bar z -H_B\right] 
\end{equation}
where by $-H_B$ we have denoted the boundary term in
eq.(\ref{azione}). 
With regard to the kinetic terms we notice that 
\begin{equation}
\dot q^c_n-\xi(q^c_n) =F_*\dot q_n
\end{equation}
if $\xi(z)$ generates the diffeomorphism $F$ and recalling
eq.(\ref{Pntn}) we have
$$
\sum_{n=1}^{\cal N} P_{ni} \dot q^i_n = \sum_{n=1}^{\cal N}
\left(F_*P_n\right)_z  \left(F_* \dot q_n\right)^z
+\left(F_*P_n\right)_{\bar z}  \left(F_* \dot q_n\right)^{\bar z}
$$
\begin{equation}\label{Pdotq}
= \sum_{n=2}^{\cal N} t_n \left(\dot z^c_n -\dot z^c_1 - \xi^z(q^c_n)+
\xi^z(q^c_1)\right) + c.c. 
\end{equation}
Then we notice that
$$
\int \pi^{ij}\dot h_{ij} \frac{i}{2}dz\wedge d \bar z=\int
F^*(\pi)^{ij}\frac{d}{dt}(F^*(\delta_{ij}e^{2\sigma}))\frac{i}{2}dz\wedge
d \bar z= 
$$
$$
=\int F^*\left(-\frac{\sqrt{h}}{4\pi}\sum_{k=2}^{\cal
N}(Q_k t_k+\bar Q_k\bar t_k)
+\sqrt{h} ~P(Y^0)\right)^{ij}~[F^*(\delta_{ij}\frac{d}{dt}e^{2\sigma})
+F^*({\cal L}_{\xi}(e^{2\sigma}\delta_{ij}))]\frac{i}{2}dz\wedge d \bar z=    
$$
\begin{equation}\label{pidoth}
=\int \left(-\frac{\sqrt{h}}{4\pi}\sum_{k=2}^{\cal
N}(Q_k t_k+\bar Q_k\bar
t_k)\right)^{ij}P(\xi)_{ij}~\frac{i}{2}dz\wedge 
d \bar z
\end{equation}
where we took into account that
\begin{equation}
\frac{d}{dt}F(t)^*[A(t)]=F(t)^*[\dot A(t)]+F(t)^*{\cal L}_\xi [A(t)].
\end{equation}
Explicit evaluation of eq.(\ref{pidoth}) gives
\begin{equation}
-\frac{1}{\pi}\int\sum_{n=2}^{\cal N} Q_{nzz} t_n
 \frac{\partial}{\partial \bar z} \xi^z \frac{i}{2} dz\wedge d\bar
 z+c.c. = \sum_{n=2}^{\cal N} t_n(\xi^z(q^c_n)-\xi^z(q^c_1))+c.c.
\end{equation}
Then summing  eq.(\ref{Pdotq}) to eq.(\ref{pidoth}) we obtain
\begin{equation}
\sum_{n=1}^{\cal N} P_{ni}\dot q^i_n + \int \pi^{ij}\dot h_{ij}
\frac{i}{2}dz\wedge\bar dz = \sum_{n=2}^{\cal N} t_n \dot {z'}_n^c +
\bar t_n \dot {{\bar z}'}_n^c 
\end{equation}
with ${z'}^c_n={z}^c_n-{z}^c_1$.

Thus we have reached the functional integral
\begin{equation}\label{reduced}
Z=\int \prod_{n=2}^{\cal N}D[{z'}^c_n]D[\bar{ z'}^c_n]D[t_n]D[\bar
t_n]~e^{i\int (\sum_{n=2}^{\cal N}( t_n\dot {z'}^c_n +
\bar t_n \dot{\bar z'}^c_n- H_B)dt}, 
\end{equation}
i.e. all functional determinants cancel out. The main point in
achieving such a result 
is the remark in \cite{carlip} that the expression
\begin{equation}
\int D[N^l]~e^{-i\int N^lH_l d^2z} 
\end{equation}
if we want to respect invariance under diffeomorphisms has to be
understood as
$
\displaystyle{\delta(\frac{H_i}{\sqrt h})}
$
and similarly for $N$ and $H$.
The role of the two gauge fixings $\delta(\chi^1)$ and
$\delta(\chi^2)$ is simply to lock the space diffeomorphisms; their
explicit form does not intervene in the final result. One could
approach with some advantage the functional integral along the so
called geometric procedure \cite{alvarez,polchinski,mottola}. Here one
start with the 
functional integral without introducing the gauge fixings $\chi^1$ and
$\chi^2$
\begin{equation}
Z= \int \prod_{n=1}^{\mathcal{N}}D[P_n]
D[\pi^{ij}] D[h_{ij}] D[N^i] D[N] \delta(\chi)
|{\rm Det}\{\chi, H\}| e^{i S},
\end{equation}
with a scalar $\chi$. We proceed exactly as before reaching instead of
eq.(\ref{functional_sigma})
\begin{equation}
\int D[F^0]~\times~A
\end{equation}
where $A$ is the diffeomorphism invariant quantity 
\begin{equation} 
\int \prod_{n=2}^{\cal N}D[{z'}^c_n]D[\bar{z'}^c_n]D[t_n]D[\bar
t_n] D[\sigma_R]~|{\rm Det}\{\chi, H\}| \delta(\frac{H}{\sqrt{h}}) 
e^{i S}.
\end{equation}
Factorising away the infinite gauge volume $\int D[F^0]$ we obtain
again eq.(\ref{reduced}). This procedure clarifies the fact that in
eq.(\ref{functional_sigma}) no trace is left of $\chi^1$ and
$\chi^2$.

\section{The boundary terms}

We come now to the boundary term 
\begin{equation}\label{bound2}
-H_B= 2 \int_{B_t} d^{(D-1)}x
\,\sqrt{\rho} N \left( K_{B_t}+\frac{\eta}{\cosh\eta}
\mathcal{D}_\alpha
v^\alpha\right) -2 \int_{B_t} d^{(D-1)}x\, r_\alpha
\pi^{\alpha\beta}_{(\rho)} N_\beta 
\end{equation}
which plays the role of the
hamiltonian. Such a term was already computed in \cite{CMS1} and found
to be given by
\begin{equation}
H_B = \ln s^2
\end{equation}
being $\ln s^2$ the constant term in the asymptotic behavior of
$2\sigma$ i.e.
\begin{equation}\label{asymptoticsigma}
2\sigma = -\mu \ln z\bar z + \ln s^2 + O(\frac{1}{z})+O(\frac{1}{\bar
z}) + O((z\bar 
z)^{\mu-1}) = -\mu \ln z\bar z + \ln s^2 + O((z \bar z)^{-\alpha}), 
\end{equation}
where $\alpha={\rm min}(1/2,1-\mu)$. 

We recall also that the
same hamiltonian $H_B$ was derived in \cite{CMS1} directly from
the equations of motion obtained by solving the constraints in the
maximally slicing gauge $K=0$.

For completeness we want to
discuss again such a term in some detail. Our treatment as the one in
\cite{carlip} is based on the ADM formalism, with the difference that
instead of having a compact space, our space (the plane) is non
compact. 

In \cite{hawkinghunter} is described the procedure  for computing the
boundary term for non compact spaces. Such a procedure amounts to
compute 
the boundary term appearing in (\ref{bound2}) on a compact region and then
letting the boundary go to infinity. In order to do so one
has to supply the fields $N$, $N^z$ and $\sigma$ for $|z|=r_0$ and the
behavior of such boundary conditions for $r_0\rightarrow \infty$.
Such asymptotic behaviors define the ADM frame at infinity. As a
rotating frame has $N^z \simeq z$ we shall impose a $N^z$ behaving at
infinity like $r^\alpha$ with $\alpha<1$. It is immediately seen
\cite{CMS1} that under such a condition the only surviving boundary
term in $H_B$ with $N$ constant on the boundary is
\begin{equation}
- 2N(r^0)\int_{Bt}  dx \sqrt{\rho} K_{Bt}
\end{equation}
whose exact expression in terms of the conformal factor is \cite{CMS1}
\begin{equation}
-2 N(r_0) \int_{Bt} d\theta  (r_0 \partial_r\sigma(r_0,\theta)+1).
\end{equation}
Thus one should proceed as follows: one solves eq.(\ref{liouville})
for $|z|\leq r_0$ with the boundary condition  
\begin{equation}\label{boundcond}
2\sigma (z) = -\mu_0\ln r^2_0 + c_0 ~~~~{\rm for}~~~~|z|=r_0.
\end{equation}
 We shall denote such solution as
$\sigma_V$. Then the boundary term on the non compact plane is given by 
the limit for $r_0\rightarrow \infty$ of
\begin{equation}\label{boundterm}
2 N(r_0) \int d\theta (-r_0 \partial_r\sigma_V(r_0,\theta)-1). 
\end{equation}
In general as pointed out in \cite{hawkinghunter} some divergence may
originate in this limiting procedure and such a divergence can be
eliminated by taking the difference between the boundary expression
and the same expression computed on some background space time. Such a
procedure can be applied to our case but, as we shall see, the divergent
term in our case is a number independent of the dynamical variables
and as such irrelevant in the hamiltonian. Thus even if such a
subtraction can be performed it is not necessary for our purposes.
Clearly solving eq.(\ref{liouville}) on the finite region $|z|\leq
r_0$ with the boundary condition (\ref{boundcond}) is far more
difficult that solving 
eq.(\ref{liouville}) on the whole plane with a given asymptotic
behavior $2\sigma=-\mu\ln z\bar z + O(1)$ as in the first case the
problem cannot be reduced to an ordinary linear fuchsian
equation. However one expects that for large $r_0$ one can replace in
the calculation of the boundary term eq.(\ref{boundterm})
$\sigma_V$ with $\sigma$ with a proper $\mu$, as for large $r_0$
eq.(\ref{asymptoticsigma}) becomes more and more circularly symmetric. To
prove this assertion
let us consider the solution of eq.(\ref{liouville}) $2\sigma(z,\mu)$
with $\mu$ defined by
\begin{equation}\label{mumu0}
-\mu\ln r^2_0 +\ln s^2({z'}^c_n, \bar {z'}^c_n,t_n,\bar t_n,\mu) 
= -\mu_0\ln r_0^2+c_0. 
\end{equation}
The difference $\eta = 2\sigma_V-2\sigma$ satisfies  
\begin{equation}\label{etaequation}
\Delta_0\eta(z) = -2 \pi^{\bar z}_z \pi^{z}_{\bar z}
e^{-2\sigma(z)}(e^{-\eta(z)}-1) = - e^{-2\tilde\sigma}(e^{-\eta(z)}-1)
\end{equation}
where for $|z|=r_0$, $\eta(z) = O(r_0^{-2\alpha})$ according to
eq.(\ref{asymptoticsigma}). Picard \cite{picard} examined
eq.(\ref{etaequation}) on a bounded domain proving that the function
$\eta$ assumes its maxima and minima on the boundary, in our case on
the circle of radius $r_0$. The proof is based on the positivity of
$e^{-2\tilde\sigma}$ and the result holds also when
$e^{-2\tilde\sigma}$ possesses locally integrable singularities
\cite{picard}, as it happens in our case. Thus we can conclude that
$\eta(z)= O(r_0^{-2\alpha})$ on the whole disk of radius $|z|=r_0$.
Integrating now eq.(\ref{etaequation}) on the disk $|z|\leq r_0$ we have
\begin{equation}
r_0\int[\partial_r\sigma_V(r_0)-\partial_r\sigma(r_0,\theta)] d \theta=
\int_{|z|\leq r_0} 
\left(- 2 \pi^{\bar z}_z \pi^{z}_{\bar z}
e^{-2\sigma(z)}(e^{-\eta(z)}-1)\right)d^2z 
=O(r_0^{-2\alpha}), 
\end{equation}
as 
\begin{equation}
\int -2\pi^{\bar z}_z\pi^z_{\bar z} e^{-2\sigma} d^2 z
\end{equation}
extended to the whole plane is convergent due to the asymptotic
behavior of $2\sigma$.
Then using eq.(\ref{mumu0}) we have
\begin{equation}
H_B =4\pi N(r_0) [\mu-1+O(r_0^{-2\alpha})].
\end{equation}
Using the fact that $\ln s^2$ is a real analytic function of $\mu$
\cite{CMS3} and 
the implicit function theorem \cite{rudin} we obtain solving
eq.(\ref{mumu0}) for large 
$r_0$ 
\begin{equation}
\mu-\mu_0 =\frac{\ln s^2({z'}^c_n,\bar{z'}^c_n, t_n,\bar t_n,\mu_0)-c_0}{\ln
r_0^2}\left(1+O(\frac{1}{\ln r^2_0})\right). 
\end{equation}
Thus 
\begin{equation}
H_B=4\pi N(r_0)\left[\frac{\ln s^2({z'}^c_n,\bar{z'}^c_n, t_n,\bar
t_n,\mu_0)-c_0}{\ln
r^2_0}\left(1+O(\frac{1}{\ln r^2_0})\right)+\mu_0-1
+O(r_0^{-2\alpha})\right] 
\end{equation}
and we have to take the limit of $H_B$ for $r_0\rightarrow \infty$. We
recall that $N(r_0)$ describes the asymptotic time and it is
immediately seen that the only choice of $N(r_0)$ which gives rise in
the limit $r_0\rightarrow \infty$, to a dynamics which is neither
singular nor frozen is 
\begin{equation}\label{Nasympt}
N(r_0)=c_1\ln r^2_0 +c_2
\end{equation}
producing  in the limit $r_0\rightarrow\infty$, with
the conventional normalization adopted in \cite{CMS1}
$c_1=\frac{1}{4\pi}$, the hamiltonian 
\begin{equation}
H_B = \ln s^2({z'}^c_n,\bar{z'}^c_n, t_n,\bar t_n,\mu_0) + {\rm const.}
\end{equation}
where the divergent numerical constant $( -c_0 +
(\mu_0-1)\ln r_0^2)$, is irrelevant to the hamiltonian. Subtracting to
the boundary terms the contribution due to the background we would
have obtained
\begin{equation}
H_B = \ln s^2({z'}^c_n,\bar{z'}^c_n, t_n,\bar t_n,\mu_0) - c_0
\end{equation}
with no divergent term. 
The non trivial part in $H_B$ is contained in $\ln s^2$ which is a
function of the canonical variables ${z'}_n^c$ and $t_n$. 
We notice that the behavior (\ref{Nasympt}) is the asymptotic
behavior of $N$ 
obtained from the solution of the classical equation of motion in
the $K=0$ gauge \cite{CMS1}. Thus we reproduced for the functional
integral the 
same expression which would have been derived from the reduced
particle dynamics i.e.
\begin{equation}\label{reduced2}
Z=\int \prod_{n=2}^{\cal N}D[{q'}^c_n]D[t_n]~e^{i\int (\sum_{n=2}^{\cal
N}(t_n\dot {z'}^c_n + \bar t_n \dot{\bar z'}^c_n) - 
\ln s^2({z'}^c_n,\bar{z'}^c_n, t_n,\bar t_n,\mu_0))dt}.  
\end{equation}
In principle one would have expected measure terms
i.e. quantum corrections to such naive translation, but as we saw,
detailed treatment starting from formula (\ref{functional1}) shows
that all the 
intervening measure terms (determinants) cancel out exactly.

Obviously the fact that we have reached expressions
(\ref{reduced},\ref{reduced2}) for the 
functional integral tells us little about the ordering
problem. In \cite{CMS1} we gave a detailed quantum treatment of the
two particle problem.  
The choice performed in \cite{CMS1} was dictated by naturalness and
aesthetic reasons reaching the logarithm of the Laplace-Beltrami
operator on a cone. As discussed in \cite{CMS1} this is very similar
to the quantum treatment of a test particle on a cone given in
\cite{deserjackiwcmp}. But there is no a priori reason for that choice. A
standard choice for the functional integral is the mid point rule
\cite{lee} which is equivalent to the Weyl ordering at the operator
level. In our case
\begin{equation}
H_B=\ln \left[(q\bar q)^{\mu_0} P\bar P\right] = \ln
\left[(q_1^2+q_2^2)^{\mu_0}(P_1^2+P_2^2)\right]  
\end{equation}
and the Weyl ordering gives rise simply to the operator 
\begin{equation}
\mu_0\ln(\hat{q}_1^2+\hat{q}_2^2)+\ln(\hat{P}_1^2+\hat{P}_2^2).
\end{equation}
This is the choice examined by Ciafaloni and Munier in
\cite{ciafaloni} in the context of high energy behavior of Yang-Mills
field theory. The time evolution operator induced by any hamiltonian
can be described by a functional integral provided the classical
hamiltonian is defined through a special ordering procedure
\cite{weinberg}. It appears that the logarithm of the Laplace-Beltrami
operator can be obtained only through a rather complicated ordering
process and the same can be said for the functional translation of the
Maass laplacian adopted in \cite{hosoya,carlip2}. For more than two
particles the hamiltonian becomes very complicated and here up to now
no guiding principle has emerged for addressing the ordering
problem. It is not clear on the other hand whether the functional or
the operator approach should be the guiding principles in quantum
gravity. Other issues common to the problem treated in \cite{carlip}
are the relations to the proper time quantization
\cite{dasgupta,ambjorn} and to the causal quantization of gravity
\cite{teitelboim}. E.g. if $N$ is integrated only on positive values
one does not obtain the strict constraint $\delta( H/\sqrt{h})$ but
only a kind of smeared form of it and one does not expect equivalence
with the functional integral (\ref{reduced2}).

\section*{Acknowledgments}

We thank G. Morchio for a useful discussion.

\section*{Appendix}

Here we show that for a proper choice of the $\xi^k$, always
respecting eq.(\ref{xipunct}), a square integrable $\xi^{0k}$
exists satisfying eq.(\ref{decomposition}). Such an equation is solved by 
\begin{equation}
\bar\xi^{0k\bar z } (z) =  \bar\xi^{k\bar z} (z) -4\Delta_0^{-1}
\left( \partial_{\bar z}(e^{-2\sigma} \sum_{k=2}^{\cal N} \beta^k_m
Q_{mzz})\right)(z) +c_k   
\end{equation}  
where $\Delta_0^{-1} = \frac{1}{4\pi}\ln(z-z')(\bar z-\bar z')$.
It is easily seen that 
\begin{equation}
\int \partial_{\bar z}(e^{-2\sigma} \sum_{k=2}^{\cal N} \beta^k_m
Q_{mzz})\frac{i}{2} dz\wedge d\bar z =0
\end{equation}  
and thus
\begin{equation}
-4\Delta_0^{-1}
\left( \partial_{\bar z}(e^{-2\sigma} \sum_{k=2}^{\cal N} \beta^k_m
Q_{mzz})\right)(z)
\end{equation}  
goes to a constant at infinity. Taking the scalar product of
eq.(\ref{decomposition}) with $\bar Q_k$ we have
\begin{equation}
\bar\xi^{0k\bar z } (z_n) -\bar\xi^{0k\bar z } (z_1) =0
\end{equation}  
and thus using the freedom on the $c_k$ we can have $\xi^{0k\bar z}
(z_n)=0$ for all $z_n$ and $k$. Now by properly choosing $\xi^{k\bar
z}$ outside a circle of radius $R$ containing all singularities,
always respecting conditions  
(\ref{xipunct}), we obtain $\bar \xi^{0k\bar z} =0$ outside such a circle.


\begin{thebibliography}{}
 
\bibitem{carlip} S. Carlip, Class. Quant. Grav. 12 (1995) 2201.

\bibitem{moncrief} V. Moncrief, J. Math. Phys. 30 (1989) 2907.

\bibitem{hosoya}  A. Hosoya, K. Nakao, Prog. Theor. Phys. 84 (1990) 739.

\bibitem{carlip2}  S. Carlip, Phys. Rev. D 42 (1990) 2647;
Phys. Rev. D 45 (1992) 3584.

\bibitem{fayterras} J.D. Fay, J. Reine Angew. Math. 293 (1977) 143;
A. Terras, ``Harmonic analysis on symmetric spaces and applications'',
Springer- Verlag, Berlin, (1985).

\bibitem{BCVW} A. Bellini, M. Ciafaloni, P. Valtancoli, Physics
Lett. B 357 (1995) 532; Nucl. Phys. B 454 (1995) 449; Nucl. Phys. B
462 (1996) 453; M. Welling, Class. Quantum Grav. 13
(1996) 653; Nucl. Phys. B 515 (1998) 436.  

\bibitem{MS} P. Menotti, D. Seminara, Ann. Phys. 279 (2000) 282;
Nucl. Phys. (Proc. Suppl.) 88 (2000) 132.

\bibitem{CMS1} L. Cantini, P. Menotti, D. Seminara, Class. Quant. Grav. 18
(2001) 2253.
 
\bibitem{CMS3} L. Cantini, P. Menotti, D. Seminara, Nucl. Phys. B 638
(2002) 351.
 
\bibitem{hawkinghunter} S.W. Hawking and C. J. Hunter, Class. Quantum
Grav. 13 (1996) 2735.

\bibitem{hayward} G. Hayward, Phys. Rev. D 47 (1993) 3275; J.D. Brown,
S.R. Lau, J.W. York; gr-qc/0010024.

\bibitem{alvarez}  O. Alvarez, Nucl. Phys. B 216 (1983) 125. 

\bibitem{polchinski}  J. Polchinski, Comm. Math. Phys. 104 (1986) 37.

\bibitem{mottola} Z. Bern, E. Mottola, S. K. Blau,
Phys. Rev. D 43 (1991) 1212.

\bibitem{bers}  L. Bers, Am. Math. Soc. 5 (1981) 131.

\bibitem{dhoker} E. D'Hoker, S. B. Giddings, Nucl. Phys. B 291 (1987)
90; E. D'Hoker, D. H. Phong, Rev. Mod. Phys. 60 (1988) 917. 

\bibitem{deser} S. Deser, Class. Quantum Gravity 2 (1985) 489.

\bibitem{polyakov} A. M. Polyakov, Phys. Lett. B 103 (1981) 207.

\bibitem{MS1} P. Menotti, D. Seminara, Nucl. Phys. Proc. Suppl. 88
(2000) 132.

\bibitem{menottipeirano}  P. Menotti, P.P. Peirano,
Nucl.Phys. B488 (1997) 719.

\bibitem{ADM} R. Arnowitt, S. Deser and C.W. Misner, in ``Gravitation:
an introduction to current research'' Edited by L.Witten, John Wiley
\& Sons New York, London 1962.
 
\bibitem{picard} E. Picard, Compt. Rend. 116 (1893) 1015;
J. Math. Pures Appl. 4 (1893) 273 and (1898) 313;
Bull. Sci. math. XXIV 1 (1900) 196. 

\bibitem{licht} L. Lichtenstein, Acta Mathematica 40 (1915) 1.
 
\bibitem{troyanov} M. Troyanov, Trans. Am. Math. Soc. 324 (1991) 793. 

\bibitem{DJH} A. Staruszkiewicz, Acta Phys. Polonica 24 (1963)
734; S. Deser, R. Jackiw and G. 't Hooft, Ann. Phys. (NY) 152
(1984) 220.

\bibitem{deserjackiw} S. Deser and R. Jackiw, Ann. Phys. 153 (1984)
405.

\bibitem{hooft} G. 't Hooft, Class. Quantum Grav. 9 (1992) 1335;
Class. Quantum Grav. 10 (1993) 1023 

\bibitem{hooft2} G. 't Hooft, Class. Quantum Grav. 10 (1993) 1653;
Class. Quantum Grav. 13 (1996) 1023. 


\bibitem{wald} R. M. Wald, ``General Relativity'', The University of
Chicago Press, Chicago and London (1984). 

\bibitem{EIH} A. Einstein, L. Infeld and B. Hoffmann, Ann. Math. 39
(1938) 65; J. N. Goldberg, in ``Gravitation: an introduction to current
research'' Edited by L.Witten, John Wiley \& Sons New York, London
1962.      

\bibitem{lee} T. D. Lee, ``Particle physics and introduction to field
theory'', Harwood  Academic Publishers, Switzerland (1988).

\bibitem{ciafaloni} M. Ciafaloni, S. Munier, e-Print
Archive: hep-th/0206106.

\bibitem{rudin} W. Rudin, ``Principles of Mathematical Analysis'',
McGraw-Hill, New York (1976).

\bibitem{henneaux} M. Henneaux, C. Teitelboim, ``Quantization of gauge
systems'', Princeton University Press, Princeton (1992).

\bibitem{deserjackiwcmp} S. Deser, R. Jackiw, Comm. Math. Phys. 118
(1988) 495.   

\bibitem{weinberg} S. Weinberg, ``The Quantum Theory of Fields'', Vol.
1, Cambridge University Press, 1995.

\bibitem{dasgupta} A. Dasgupta, R. Loll, Nucl. Phys. B 606 (2001) 357.

\bibitem{ambjorn} J. Ambjorn, J. Jurkiewicz, R. Loll,
Phys. Rev. Lett. 85 (2000) 924; Nucl. Phys. B 610 (2001) 347;
J. Ambjorn, e-Print Archive: gr-qc/0201028 and references therein.

\bibitem{teitelboim} C. Teitelboim, Phys. Rev. D 25 (1982) 3159;
M. Henneaux, C. Teitelboim, J. D. Vergara, Nucl. Phys. B 387 (1992)
391.

\end{thebibliography}
\end{document}